\documentclass[twocolumn,preprintnumbers,amsmath,amssymb,aps,prl,reprint]{revtex4}
\usepackage{graphicx}
\begin{document}

\title{
Controlled Fluidization, Mobility and Clogging in Obstacle Arrays Using Periodic Perturbations
} 
\author{
C. Reichhardt and C.J.O. Reichhardt 
} 
\affiliation{
Theoretical Division and Center for Nonlinear Studies,
Los Alamos National Laboratory, Los Alamos, New Mexico 87545, USA
} 

\date{\today}
\begin{abstract}
We show that the clogging susceptibility and flow 
of particles moving through a random obstacle array 
can be controlled with a transverse or longitudinal ac drive.
The flow rate can vary
over several orders of magnitude, and we find
both an optimal frequency and
an optimal amplitude of driving that maximizes the flow.
For dense arrays, 
at low ac frequencies a heterogeneous creeping clogged phase appears
in which rearrangements between different clogged configurations occur.
At intermediate frequencies 
a high mobility fluidized state forms, and 
at high frequencies the system reenters 
a heterogeneous frozen clogged state. 
These results provide a technique for
optimizing flow through 
heterogeneous media that could also
serve as the basis for a particle separation method. 
\end{abstract}
\maketitle

Particle transport through heterogeneous media 
is relevant to flows in porous media \cite{1,2}, 
transport of colloidal particles on ordered or disordered substrates \cite{3,4,5,6,7},
clogging phenomena \cite{8,9,10,11,12,13}, filtration \cite{14,15,16}, 
and active matter motion in disordered environments \cite{17,18,19,20}.
It also has similarities to systems that exhibit depinning phenomena 
when driven over random or ordered substrates \cite{21}.
Recent
work
has focused on
clogging effects
for particle motion through
obstacle arrays, where the onset of clogging 
is characterized by the formation of
a heterogeneously dense state \cite{11,12,13}. 
Such clogging is relevant for the performance of filters
or for limiting the amount of flow through disordered media, 
so understanding how to avoid clog formation or
how to optimize the particle mobility in
obstacle arrays is highly desirable. 
Clogging also occurs for particle flow through hoppers or constrictions, 
where there can be a transition from a flowing to a clogged state
as the aperture size decreases or the flow rate increases
\cite{22,23,24,25,26}.
The clogging susceptibility in such systems can be reduced
with periodic perturbations or vibrations \cite{27,28,29}.
Applied perturbations generally produce
enhanced flows in disordered systems \cite{30,31,N1,32,33};
however, there are examples
where
the addition of perturbations or noise can 
decrease the flow or induce jamming,
such as the freezing by heating phenomenon
\cite{34,35} or the appearance of a reentrant high viscosity state in 
vibrated granular matter \cite{36}.
A natural question is whether clogging and mobility  
for particle flows through obstacles can be
controlled or optimized with applied perturbations in the same way as
hopper flow.
The situation is more complex
for two-dimensional (2D) disordered
obstacle arrays than for hopper geometries since shaking can be applied
in either the longitudinal or transverse direction,
and one type of shaking may be more effective than the other.   

In this work we numerically examine 
particle flow though a disordered obstacle array 
where the particles experience both a dc drive and ac shaking.   
In the absence of the ac shaking,
there is a well defined clogging transition at a critical obstacle density $\phi_c^{dc}$
above which the flux of particles drops to zero.
We find that application of a transverse or longitudinal 
ac drive above
$\phi_c^{dc}$
unclogs the system and permits flow to occur,
while the mobility drops back to zero at
a higher second critical obstacle density $\phi_c^{ac}$.
We identify an optimal ac frequency for mobility
and find that at low frequencies,
the system forms a nearly immobile
heterogeneous creeping clogged state
in which particle rearrangements produce transitions between
different clogged configurations.
At intermediate frequencies, a more uniform 
flowing fluidized state appears, and
at high frequencies a heterogeneous frozen clogged state
emerges in which there are no particle
rearrangements.
The mobility for fixed frequency and changing ac amplitude is
also nonmonotonic.
For obstacle densities below $\phi_{c}^{dc}$,
the ac drive still strongly affects 
the flow rate,
and we find an optimal frequency that maximizes the flow
as well as a
local minimum in the mobility
produced by a resonance effect of the ac motion
with the average spacing between obstacles.
In most cases, transverse ac drives
produce higher mobility than longitudinal ac drives; however,
at low obstacle densities the transverse ac drive reduces the flow.
We show that these effects are robust for
a wide range of particle densities, and we map a
dynamic phase digram describing the
fluid regime, the creeping clogged phase, and the frozen clogged state.

{\it Simulation and System---} 
We simulate a 2D system of non-overlapping 
repulsive particles in the form of disks interacting 
with a random array of obstacles
where the particles are subjected to a dc drift force and an ac shaking force. 
The sample is of size $L \times L$ with $L = 100$
and we impose periodic boundary conditions in the $x$ and $y$ directions.
Interactions between pairs of disks $i$ and $j$ are given by the
repulsive harmonic force
${\bf F}^{ij}_{dd} = k(r_{ij} -2R_{d})\Theta(r_{ij} -2R_{d}){\hat {\bf r}}_{ij}$, where
the disk radius $R_d=0.5$,
$r_{ij} = |{\bf r}_{i} - {\bf r}_{j}|$, 
$ {\hat {\bf r}}_{ij} = ({\bf r}_{i} - {\bf r}_{j})/r_{ij}$,  and $\Theta$ is the Heaviside step
function. 
The spring stiffness 
$k = 200$ is large enough that disks overlap by less than one percent, placing
us in the hard disk limit as confirmed by
previous works \cite{11,12,37}. 
The obstacles are modeled as
immobile disks with the same radius and disk-disk interactions as the
mobile particles.
There are $N_m$ mobile particles with an area coverage of 
$\phi_{m} = N_{m}\pi R^{2}_{d}/L^2$,
while the area coverage of the $N_{\rm obs}$ obstacles 
is $\phi_{\rm obs} = N_{\rm obs}\pi R^2_{d}/L^2$
and the total area coverage is
$\phi_{\rm tot} = \phi_{m} + \phi_{\rm obs}$.
For monodisperse disks the
system forms a triangular solid at $\phi_{\rm tot} = 0.9$ \cite{37}.
The obstacles are placed in a dense lattice and 
randomly diluted until the desired $\phi_{obs}$ is reached,
so that the minimum spacing between obstacle centers is $d_{\rm min}=2.0$.
The particle dynamics are governed by the following overdamped equation of motion:
$\eta d {\bf r}_i/dt = {\bf F}_{\rm inter}^i + {\bf F}_{\rm obs}^i + {\bf F}_{dc} + {\bf F}_{ac}$.
Here ${\bf F}_{\rm inter}^i=\sum_{j=0}^{N_m}{\bf F}_{dd}^{ij}$ are the particle-particle
interactions, ${\bf F}_{\rm obs}^i=\sum_{k=0}^{N_{\rm obs}}{\bf F}_{dd}^{ik}$ are
  the particle-obstacle interactions, 
and  ${\bf F}_{dc} = F_{dc}{\hat {\bf x}}$ is the dc drift force applied in the
positive $x$-direction, where $F_{dc}=0.05$.
Each simulation time step is of size $dt=0.002$.
  We apply a sinusoidal ac drive that is either transverse (perpendicular) to
  the dc drive, ${\bf F}_{ac}=F_{ac}^{\perp}{\hat {\bf y}}$,
  or longitudinal (parallel) to the dc drive,
  ${\bf F}_{ac}=F_{ac}^{||}{\hat {\bf x}}$ .
  We measure
  the time average of the velocity per particle in the dc drift direction,
  $\langle V_{x}\rangle = N_m^{-1}\sum^{N_m}_{i =1}{\bf v}_{i}\cdot {\hat {\bf x}}$,
  where ${\bf v}_i$ is the velocity of particle $i$.  We define the mobility as
 $M = \langle V_{x}\rangle/ \langle V^{0}_{x}\rangle $,
  where $\langle V^{0}_{x}\rangle$ is the obstacle-free drift velocity,
  so that in the free flow limit, $M = 1.0$.
  We wait at least $10^7$ simulation time steps before taking measurements
  to ensure that the system has reached a steady state.

\begin{figure}
\includegraphics[width=\columnwidth]{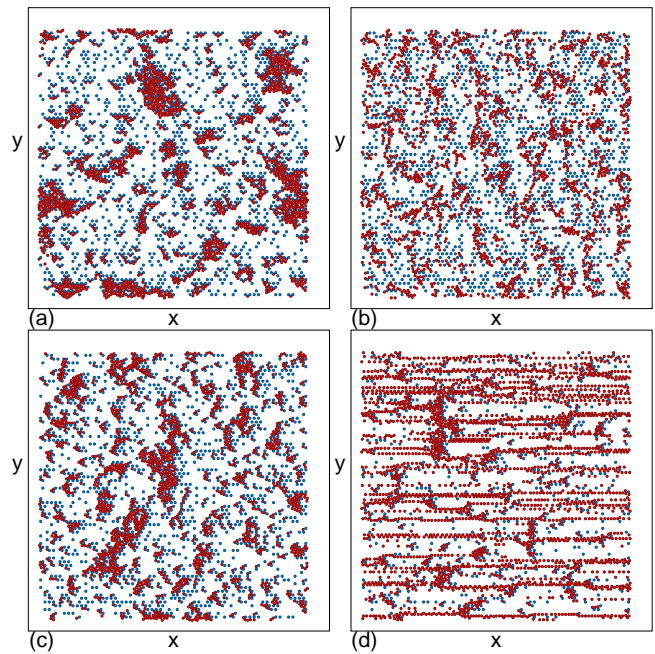}
\caption{Locations of particles (red) and obstacles (blue)
  for a system with $\phi_{\rm tot}=0.275$ and
  an $x$-direction drift force of $F_{dc} = 0.05$
  under an applied transverse ($y$-direction) ac drive of magnitude
  $F^{\perp}_{ac} = 0.5$ for different ac frequencies $\omega$.
  (a) A low mobility creeping clogged state with $M=0.01$
  at $\omega = 10^{-7}$ and $\phi_{\rm obs}=0.1256$.
  (b) A high mobility fluidized state
  with $M=0.27$ 
  at $\omega = 10^{-4}$ and $\phi_{\rm obs}=0.1256$.
  (c) A frozen clogged state with $M=0$ at
  $\omega = 10^{-1}$ and $\phi_{\rm obs}=0.1256$.
(d) A flowing state at $\omega = 10^{-1}$ and $\phi_{\rm obs} = 0.047$.
  The images in (a,b,c) were obtained at the points marked
  a, b, and c in Fig.~\ref{fig:2}(b).
}
\label{fig:1}
\end{figure}

{\it Results --}
In Fig.~\ref{fig:1}(a) we illustrate the positions of the particles and obstacles in
a sample
with $F_{dc} = 0.05$, $\phi_{\rm tot}=0.275$, and $\phi_{\rm obs}=0.1256$ under a
transverse drive of magnitude
$F^{\perp}_{ac} = 0.5$ in what we define as the low frequency limit of
$\omega = 10^{-7}$, where the mobility is very small, $M=0.01$.
The particles assemble into high density clogged regions separated by large void areas.
There are slow rearrangements of the particles but little net motion in the direction of
the dc drift,
so the system is effectively transitioning between different clogged configurations. 
At $\omega=10^{-4}$ in
Fig.~\ref{fig:1}(b),
the mobility of the same sample reaches its maximum value of
$M = 0.27$.
Here the clustering is reduced compared to what occurs at lower frequencies,
and the system is in a partially fluidized state. 
For the high frequency of $\omega=10^{-1}$ in 
Fig.~\ref{fig:1}(c),
a completely frozen clogged state with $M = 0$ appears.
In Fig.~\ref{fig:1}(d), when $\omega=10^{-1}$ but the obstacle density is reduced
to $\phi_{\rm obs}=0.047$, 
the system is in a flowing state.

\begin{figure}
  \includegraphics[width=\columnwidth]{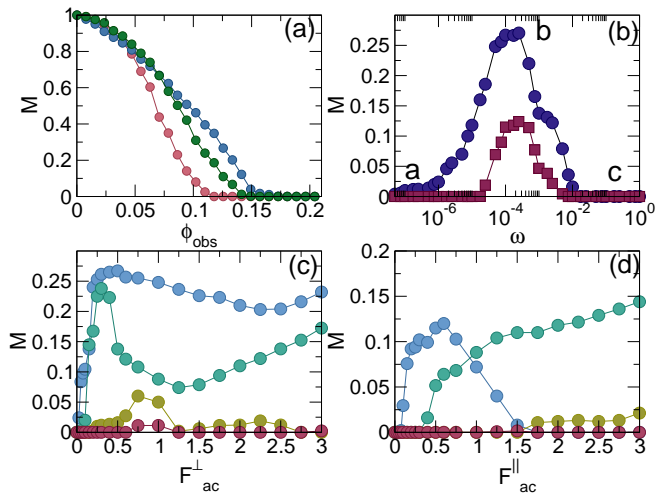}
  \caption{(a) Mobility $M$ vs obstacle density $\phi_{\rm obs}$
    for $\phi_{\rm tot} = 0.275$ at $F^{\perp}_{ac} = F^{||}_{ac}=0$ (pink), where
    $M = 0$ for $\phi_{\rm obs} > 0.115$;
    at $F^{\perp}_{ac} = 0.5$ and $\omega = 10^{-4}$ (blue),
    where $M \approx 0$ for $\phi_{\rm obs} > 0.195$;
    and at $F^{||}_{ac} = 0.5$ and $\omega = 10^{-4}$ (green),
    where $M \approx 0$ for $\phi_{\rm obs}>0.155$. 
    (b) $M$ vs ac frequency $\omega$
    for the system in Fig.~\ref{fig:1}(a--c) at
    $\phi_{\rm tot} = 0.275$, $\phi_{\rm obs} = 0.1256$, and $F_{ac}= 0.5$
    for transverse (blue circles) and longitudinal (red squares) ac driving
    showing a low frequency clogged state,
    an intermediate frequency flowing state, and a high frequency clogged state.
    The letters a, b, c mark the frequencies at which the images in Fig.~\ref{fig:1}(a--c) were
    obtained.
    (c) $M$ vs $F^{\perp}_{ac}$ for the system in (b)
    under transverse driving with $\omega = 10^{-4}$ (blue), 
    $10^{-3}$ (green), $10^{-2}$ (gold), and $10^{-1}$ (red).
    (d) M vs $F_{ac}^{||}$ at the same frequencies as
    in (c) under longitudinal driving.
}
\label{fig:2}
\end{figure}

In Fig.~\ref{fig:2}(a) we plot $M$ versus obstacle density $\phi_{\rm obs}$ for 
a system with $\phi_{\rm tot} = 0.275$ for zero ac drive,
a transverse ac drive of $F^{\perp}_{ac} = 0.5$ at $\omega = 10^{-4}$,
and a transverse dc drive with $F^{||}_{ac}=0.5$ and $\omega=10^{-4}$.
A clogged state with $M=0$ appears for $\phi_{\rm tot}>0.115$ under no ac drive,
for $\phi_{\rm tot}>0.2$ under transverse ac driving,
and for $\phi_{\rm tot}>0.155$ under longitudinal driving,
so there is a wide range of frequencies over which
the transverse ac drive is the most effective at reducing clogging.  
For $\phi_{\rm obs} < 0.07$, the transverse ac drive produces
a lower mobility $M$ than either the
longitudinal or zero ac driving.

In Fig.~\ref{fig:2}(b) we plot $M$ versus
ac frequency $\omega$ for the system
from Fig.~\ref{fig:1}(a--c) with $\phi_{\rm tot} = 0.275$
and $\phi_{\rm obs} = 0.1256$
for transverse and longitudinal ac driving of magnitude $F_{ac}=0.5$.
We find a low mobility state for
$\omega < 10^{-6}$  and a zero mobility state for $\omega \geq 10^{-2}$.
The optimal mobility occurs at a frequency of
$\omega \approx 2.5 \times 10^{-4}$.
Both directions of ac driving produce the same dynamic states,
but the maximum value of $M$ for longitudinal driving
is less than half that found for transverse driving,
and the window of unclogged states is narrower for longitudinal driving.
Additionally,
the low frequency states with $\omega < 10^{-5}$
are fully clogged with $M = 0$ for longitudinal driving, but have a small finite
mobility for transverse driving.
These results indicate that there are
two different types of clogged states separated by an
intermediate fluidized state in which the mobility reaches its optimum value.

In Fig.~\ref{fig:2}(c) we plot $M$ versus $F^{\perp}_{ac}$ for a system with
$\phi_{\rm tot} = 0.275$ and $\phi_{\rm obs} = 0.1256$ at the optimal 
frequency of $\omega=10^{-4}$ and at
$\omega=10^{-3}$, $10^{-2}$, and $10^{-1}$.  For each driving frequency, there is
an optimal value of $F^{\perp}_{ac}$ that maximizes $M$.
Figure~\ref{fig:2}(d) shows $M$ versus $F^{||}_{ac}$ for the same system at the
same driving frequencies.
At $\omega = 10^{-4}$,
$M$ initially increases with $F^{||}_{ac}$ but it then decreases
until the system reaches a clogged state with
$M = 0$ for $F^{||}_{ac} > 1.5$.
Previous studies of
particles moving over randomly placed obstacles under a purely dc drive
have shown that negative differential conductivity or a zero mobility
state can appear at high dc drives
\cite{38,39,40,41}.
In our system we find a similar effect under large
longitudinal ac drives,
so that in general the system reaches a clogged state for high $F^{||}_{ac}$.
For 
$\omega = 10^{-3}$ in Fig.~\ref{fig:2}(d), $M$ increases monotonically over
the range of $F^{||}_{ac}$ shown;
however, $M$ does decrease for much larger values of $F^{||}_{ac}$.
In general,
$M$ is higher for transverse ac driving since the transverse shaking permits the particles
to more easily move around obstacles, whereas for longitudinal ac driving,
the particles are pushed toward the obstacles and $M$ is reduced.

\begin{figure}
\includegraphics[width=\columnwidth]{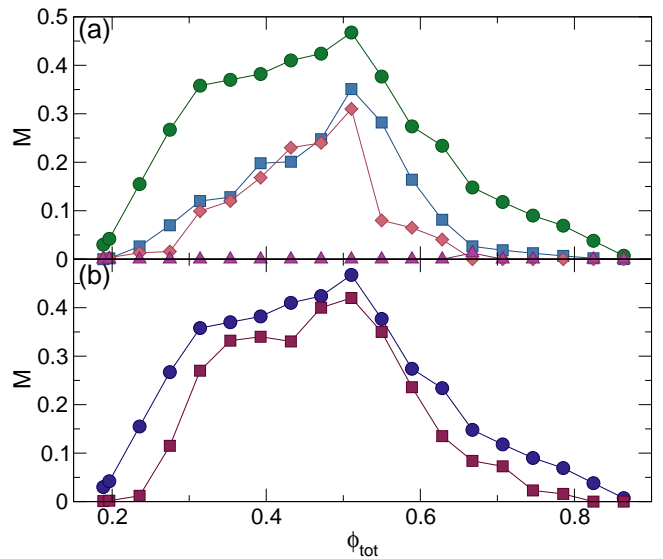}
\caption{(a) $M$ vs $\phi_{\rm tot}$ for $\phi_{\rm obs} = 0.1256$
  and $F^{\perp}_{ac} = 0.5$ at $\omega = 5.0\times 10^{-6}$ (black squares),
  $10^{-4}$ (red circles), $10^{-2}$ (green diamonds) and $10^{-1}$ (blue triangles).
  Over the entire range of $\phi_{\rm tot}$, the $\omega=10^{-4}$ curve has
  the highest values of $M$.
  (b) $M$ vs $\phi_{\rm tot}$ in the same system
  for transverse (blue circles) and longitudinal (red squares) ac driving 
  at $\omega=10^{-4}$, where transverse ac driving produces the highest values of $M$.
}
\label{fig:3}
\end{figure}

In Fig.~\ref{fig:3}(a) we plot $M$ versus $\phi_{\rm tot}$ for
samples with $\phi_{\rm obs} = 0.1256$ and $F^{\perp}_{ac} = 0.5$ at
$\omega = 5.0\times 10^{-6}$, $10^{-4}$,
$10^{-2}$, and $10^{-1}$.
$M$ is always small at low
$\phi_{\rm tot}$, increases to a local maximum  
at $\phi_{\rm tot} = 0.5$,
and decreases to zero as $\phi_{\rm tot}$ approaches $\phi_{\rm tot}=0.85$,
corresponding to the density at which the 
system starts to form a crystallized solid state \cite{37,42}.
We find the highest mobility for
$\omega = 10^{-4}$,
particularly for $0.66 < \phi_{\rm tot} < 0.85$
where $M$ is close to zero for $\omega = 10^{-2}$ and $10^{-1}$.
In Fig.~\ref{fig:3}(b) we show $M$ versus $\phi_{\rm tot}$
at $\omega = 10^{-4}$ for transverse and longitudinal ac driving,
where we again find that the transverse ac driving gives
higher values of $M$ for all $\phi_{\rm tot}$ and where
the local maximum in $M$ falls at $\phi_{\rm tot} = 0.5$ for both ac driving directions.

\begin{figure}
\includegraphics[width=\columnwidth]{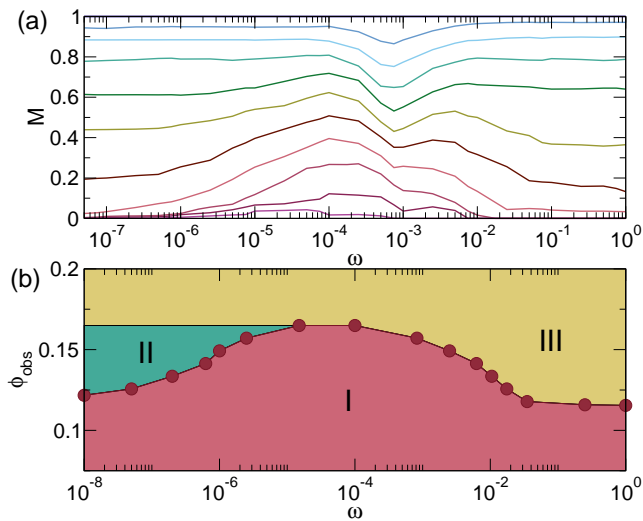}
\caption{ (a) $M$ vs $\omega$ for samples with $\phi_{\rm tot} = 0.275$
  and $F^{\perp}_{ac} = 0.5$ at
  $\phi_{\rm obs} = 0.00157$, 0.031416, 0.047124, 0.062831,
  0.07754, 0.09424, 0.1099, 0.12566, 
  0.14137, and $0.157$, from top to bottom.
  (b) Dynamic phase diagram as a function of $\phi_{\rm obs}$ vs $\omega$
  for transverse driving with $F^{\perp}_{ac}=0.5$. I:
 flowing fluidized state; II: creeping clogged state; III: frozen clogged state. 
}
\label{fig:4}
\end{figure}

In Fig.~\ref{fig:4}(a) we plot $M$ versus $\omega$ in samples with
$\phi_{\rm tot}=0.275$ and $F^{\perp}_{ac}=0.5$ 
at $\phi_{\rm obs} = 0.00157$
to $0.157$.
For $\phi_{\rm obs} > 0.1099$, the system reaches a fully clogged
state with $M  = 0$.
We define the onset of the low frequency clogged state as the point at which
$M < 0.02$. 
A local maximum in $M$ appears near $\omega = 2.5 \times 10^{-4}$ and
shifts to slightly lower frequencies as $\phi_{\rm obs}$ decreases.
A local minimum
near $\omega = 10^{-3}$ develops when
$\phi_{\rm obs} < 0.1099$,
and this minimum also shifts to lower frequencies with decreasing $\phi_{\rm obs}$.
Both of the local extrema are
correlated with characteristic length scales of the system.
The local maximum at $\phi_{\rm obs} = 0.1256$ falls at a value of $\omega$ for which
the distance $d_{\tau}=\omega^{-1}dt(F_{ac}^{\perp}/\sqrt{2}+F_{dc})$
a particle moves
during a single ac cycle matches
the average spacing $1/\sqrt{\phi_{\rm obs}}$ between obstacles.
As this average spacing decreases for increasing $\phi_{\rm obs}$,
the frequency at which the maximum value
of $M$ occurs decreases as well.
The frequency at which the local minimum appears for $\phi_{\rm obs}<0.1099$ 
corresponds to the point at which $d_{\tau}$
matches
the minimum transverse surface-to-surface obstacle spacing of $d_{\rm min}-2R_d$.
At this matching frequency, the particles preferentially collide with the obstacles
rather than moving between them or around them, reducing the mobility.  
The two resonant frequencies are separated by a factor of
10 since $F^{\perp}_{ac}/F_{dc} = 10$. 

In Fig.~\ref{fig:4}(b) we plot a dynamic phase diagram
as a function of $\phi_{\rm obs}$ versus $\omega$ for samples
with $F^{\perp}_{ac}=0.5$.
Here phase I is the flowing fluidized state, phase II
is the low frequency creeping clogged state, and phase III is the frozen clogged state. 
For $\phi_{\rm obs}> 0.165$, the spacing between
obstacles becomes so small that the system is in a frozen state for all values of $\omega$. 
The fluidized state is of maximum extent between
$\omega = 10^{-5}$ and $\omega=10^{-4}$.
The dynamic phase diagram for longitudinal ac driving (not shown) is similar;
however,
the extent of phase I is reduced.

Our results resemble what has been found in
recent experiments on the viscosity of vibrated granular
matter,
where the system is
in a jammed state for low vibration frequencies,
enters a low viscosity fluid state at intermediate frequencies,
and shows a reentrant jammed state at high frequencies \cite{36}.
Other studies have also revealed optimal frequencies
for dynamic resonances in granular matter,
where the speed of sound is the lowest 
at intermediate frequencies when the grains are the least jammed \cite{43}.

Our results show that the clogging and flow of particulate matter moving through
heterogeneous media can be controlled with ac driving,
which could be applied to colloidal particles moving
through disordered or porous media. 
Since the mobility
is a function of the driving frequency,
ac driving could be used
to separate different particle species when one species is in a low
mobility or clogged state for a given frequency while the other is in a
high mobility state.
These results can be generalized 
to the depinning dynamics in many other systems such
as active matter, vortices in superconductors, or frictional systems,
where there is a competition between the collective interactions of the
particles and quenched disorder in the substrate.   

{\it Summary---} 
We have examined the clogging and flow 
of particles moving through random obstacle arrays
under a dc drift
and an additional transverse or longitudinal ac drive.
At zero ac driving, there is a well defined obstacle density
above which the system reaches a clogged state.
When ac driving is added, this
clogging transition shifts to much higher obstacle densities. 
For large obstacle densities, we find 
a low frequency creeping clogged state where the particles undergo
rearrangements
from one clogged configuration to another with a drift mobility that is
nearly zero.
At intermediate frequencies, the particles form a high mobility fluidized state,
while at high frequencies, a zero mobility frozen clogged state appears,
so that there is an optimal mobility at intermediate frequencies.
The mobility is also nonmonotonic
as a function of ac driving amplitude for fixed ac driving frequency. 
In most cases the transverse ac driving is more effective at increasing the mobility
than longitudinal ac driving.
When the ac amplitude and frequency are both fixed,
we find that there is an optimal disk density that maximizes the mobility,
while for high disk densities the 
system enters a low mobility jammed state.
At low obstacle densities the system is always in
a flowing state; however, for transverse ac driving we find a resonant frequency
with reduced flow
when the
magnitude of the
transverse oscillations matches the minimum transverse spacing of the obstacles. 
We map a dynamic phase diagram showing the
locations of the flowing state, creeping clogged state,
and frozen clogged state.
Our results suggest that ac driving could be used
to avoid clogging and to optimize particle flows in disordered media,
and this technique could also be used as a  method
for separating different species of particles.
Our results can be generalized for controlling flows in a wide class of
collectively interacting particle systems in heterogeneous environments,
including colloids, bubbles,
granular matter, vortices in superconductors, and skyrmions in chiral magnets.

\begin{acknowledgments}
We gratefully acknowledge the support of the U.S. Department of
Energy through the LANL/LDRD program for this work.
This work was carried out under the auspices of the 
NNSA of the 
U.S. DoE
at 
LANL
under Contract No.
DE-AC52-06NA25396 and through the LANL/LDRD program.
\end{acknowledgments}

\end{document}